\documentstyle[preprint,aps]{revtex}
\tightenlines
\begin{document}
\title{TOTAL ENERGY OF THE BIANCHI TYPE I UNIVERSES}
\author{S.~S.~Xulu\thanks{%
E-mail: ssxulu@pan.uzulu.ac.za}}
\address{Department of Applied Mathematics, University of Zululand,\\
Private Bag X1001, 3886 Kwa-Dlangezwa, South Africa}

\maketitle
\begin{abstract}
Using the symmetric energy-momentum complexes of Landau and Lifshitz,
Papapetrou, and Weinberg we obtain the energy of the universe in anisotropic
Bianchi type I cosmological models . The energy (due to matter plus field)
is found to be zero and this agrees with a previous result of Banerjee and
Sen who investigated this problem using the Einstein energy-momentum
complex. Our result supports the importance of the energy-momentum complexes
and contradicts the prevailing ``folklore'' that different energy-momentum
complexes could give different and hence unacceptable energy distribution in
a given space-time. The result that the total energy of the universe in
these models is zero supports the viewpoint of Tryon. Rosen computed the
total energy of the closed homogeneous isotropic universe and found that to
be zero, which agrees with the studies of Tryon.
\end{abstract}

\section{Introduction}
\label{sec:intro}
The landmark observations of $2.7 K$ background radiation strongly support
that some version of the big bang theory is correct. Tryon (1973) assumed
that our Universe appeared from nowhere about $10^{10}$ years ago and
mentioned that the conventional laws of physics need not have been violated
at the time of creation of the Universe. He proposed that our Universe must
have a zero net value for all conserved quantities. He presented some
arguments indicating that the net energy of our Universe may be indeed zero.
His big bang model (in which our Universe is a fluctuation of the vacuum)
predicted a homogeneous, isotropic and closed Universe consisting of matter
and anti-matter equally. Tryon (1973) also referred to an elegant
topological argument by Bergmann that any closed universe has zero energy.

The subject of the energy of the Universe remained in almost a
``slumbering'' state for a long period of time and was re-opened by an
interesting work of Rosen (1994) and Cooperstock (1994). Rosen (1994)
considered a closed homogeneous isotropic universe described by a
Friedmann-Robertson-Walker (FRW) metric. He used the Einstein
energy-momentum complex\footnote{%
To avoid any confusion we mention that we use the term energy-momentum
complex for one which satisfies the local conservation laws and gives the
contribution from the matter (including all non-gravitational fields) as
well as the gravitational field. Rosen (1994) used the term pseudo-tensor
for this purpose. We reserve the term energy-momentum pseudotensor for the
part of the energy-momentum complex which comes due to the gravitational
field only.} and found that the total energy is zero. This interesting
result fascinated some general relativists, for instance, Johri {\em et al.}
(1995) and Banerjee and Sen (1997). Johri {\em et al.} (1995), using the
Landau and Lifshitz energy-momentum complex, showed that the total energy of
an FRW spatially closed universe is zero at all times irrespective of
equations of state of the cosmic fluid. They also showed that the total
energy enclosed within any finite volume of the spatially flat FRW universe
is zero at all times.

The Bianchi type $I$ solutions, under a special case, reduce to the
spatially flat FRW solutions. Banerjee and Sen (1997), using the Einstein
energy-momentum complex, studied the Bianchi type $I$ solutions and found
that the total (matter plus field) energy density is zero everywhere. As the
spatially flat FRW solution is a special case of the Bianchi type $I$
solutions, one observes that the energy-momentum complexes of Einstein and
Landau and Lifshitz give the same result for the spatially flat FRW
solutions.

These results though appear to be very interesting are usually not taken
seriously because the use of energy-momentum complexes are restricted to
``Cartesian'' coordinates. There are many prescriptions for obtaining energy
in a curved space-time, the uses of some of them (quasi-local masses) are
not limited to a particular coordinates system whereas some of them
(energy-momentum complexes) are restricted to the use of ``Cartesian
coordinates.'' A large number of definitions of quasi-local mass (associated
with a closed two-surface) have been proposed (see in Brown and York 1993).
Bergqvist (1992) studied several different definitions of quasi-local masses
for the Kerr and Reissner-Nordstr\"{o}m space-times and came to the
conclusion that not even two of these definitions gave the same results.
Contrary to this, in the last decade, several authors studied
energy-momentum complexes and obtained stimulating results. We will discuss
some of them in brief. The leading contributions of Virbhadra and Virbhadra
and his collaborators (Rosen, Parikh, Chamorro and Aguirregabiria) have
demonstrated with several examples that for a given space-time different
energy-momentum complexes give the same and acceptable energy distribution.
Several energy-momentum complexes have been shown to give the same energy
distribution for each of the following space-times: the Kerr-Newman
space-time (Virbhadra 1990, Cooperstock and Richardson 1992), Vaidya
space-time (Virbhadra 1992), Einstein-Rosen space-time (Virbhadra and Rosen
1993, Virbhadra 1995), Bonnor-Vaidya space-time (Chamorro and Virbhadra
1995), all Kerr-Schild class space-times (Aguirregabiria {\it {\em et al.}, }
1996). Recently Virbhadra (1999) discussed that the concept of local or
quasi-local mass are very much useful in investigating the {\em Seifert
conjecture} (Seifert 1979) and the {\em hoop conjecture} (Thorne 1972). He
also showed that, for a general nonstatic spherically symmetric space-time
of the Kerr-Schild class, the Penrose quasi-local mass definition (for the
Penrose definition see Penrose 1982) as well as several energy-momentum
complexes yield the same results. For some other interesting papers on this
subject see Virbhadra and Parikh (1993, 1994), Chamorro and Virbhadra (1996)
and Xulu (1998a,b). We have already discussed that Banerjee and Sen (1997)
studied energy distribution with Bianchi type I metrics, using the Einstein
definition. It is our present interest to investigate whether or not some
other energy-momentum complexes yield the same results for the Bianchi type
I metrics. We use the convention that Latin indices take values from $0$ to $%
3$ and Greek indices values from $1$ to $3$ , and take the geometrized units 
$G=1$ and $c=1$


\section{Bianchi type I space-times}

The Bianchi type I space-times are expressed by the line element 
\begin{equation}
ds^{2}=dt^{2}-e^{2l}dx^{2}-e^{2m}dy^{2}-e^{2n}dz^{2},  \label{eqtn1}
\end{equation}
where $l$, $m$, $n$ are functions of $t$ alone. The nonvanishing components
of the energy-momentum tensor $T_{i}^{\ k}$ ( $\equiv \frac{1}{8\pi }%
G_{i}^{\ k}$, where $G_{i}^{\ k}$ is the Einstein tensor) are 
\begin{eqnarray}
T_{0}^{\ 0} &=&\frac{1}{8\pi }\ \left( \ \dot{l}\dot{m}+\dot{m}\dot{n}+\dot{n%
}\dot{l}\ \right) ,  \nonumber \\
T_{1}^{\ 1} &=&\frac{1}{8\pi }\ \left( \ \dot{m}^{2}+\dot{n}^{2}+\dot{m}\dot{%
n}+\ddot{m}+\ddot{n}\ \right) ,  \nonumber \\
T_{2}^{\ 2} &=&\frac{1}{8\pi }\ \left( \ \dot{n}^{2}+\dot{l}^{2}+\dot{n}\dot{%
l}+\ddot{n}+\ddot{l}\ \right) ,  \nonumber \\
T_{3}^{\ 3} &=&\frac{1}{8\pi }\ \left( \ \dot{l}^{2}+\dot{m}^{2}+\dot{l}\dot{%
m}+\ddot{l}+\ddot{m}\ \right) .  \label{eqtn2}
\end{eqnarray}
The dot over $l,m,n$ stands for the derivative with respect to the
coordinate $t$. The metric given by Eq. ($\ref{eqtn1}$) reduces to the
spatially flat Friedmann-Robertson-Walker metric in a special case. With $%
l(t)=m(t)=n(t)$, defining $R\left( t\right) =e^{l\left( t\right) }$ and
transforming the line element ($\ref{eqtn1}$) to $t,x,y,z$ coordinates
according to $x=r\sin \theta \cos \phi ,\ y=r\sin \theta \sin \phi ,\
z=r\cos \theta $ gives 
\begin{equation}
ds^{2}=dt^{2}-\left[ R\left( t\right) \right] ^{2}\left\{ dr^{2}+r^{2}\left(
d\theta ^{2}+sin^{2}\theta d\phi ^{2}\right) \right\} ,  \label{eqtn3}
\end{equation}
which describes the well-known spatially flat Friedmann-Robertson-Walker
space-time. 

\section{The Landau and Lifshitz energy-momentum complex}

The symmetric energy-momentum complex of Landau and Lifshitz (1987) is given
by 
\begin{equation}
L^{ij}=\frac{1}{16\pi }{\cal S}_{\quad ,kl}^{ijkl}\text{,}  \label{eqtn4}
\end{equation}
where 
\begin{equation}
{\cal S}^{ijkl}=-g(g^{ij}g^{kl}-g^{ik}g^{jl})\text{.}  \label{eqtn5}
\end{equation}
$L^{00}$ \ and \ $L^{\alpha 0}$ \ give the energy density and energy current
(momentum) density components, respectively. The energy-momentum complex of
Landau and Lifshitz (LL) satisfies the local conservation laws 
\begin{equation}
\frac{\partial L^{ik}}{\partial x^{k}}=0\text{,}  \label{eqtn6}
\end{equation}
where 
\begin{equation}
L^{ik}=-g\left( T^{ik}+t^{ik}\right) \text{.}  \label{eqtn7}
\end{equation}
$g$ is the determinant of the metric tensor $g_{ik}$, $T^{ik}$ is the
energy-momentum tensor of the matter and all non-gravitational fields, and $%
t^{ik}$ is known as LL energy-momentum pseudotensor. Thus the locally
conserved quantity $L^{ik}$ contains contributions from the matter,
non-gravitational and gravitational fields. For the expression for $t^{ik}$
see Landau and Lifshitz (1987).

Integrating $L^{ik}$ over the three-space gives the energy and momentum
components 
\begin{equation}
P^{i}=\int \int \int L^{i0}dx^{1}dx^{2}dx^{3}  \label{eqtn8}
\end{equation}
$P^{0}$ is the energy and $P^{\alpha }$ are momentum components. In order to
calculate the energy and momentum density components of the line element $(%
\ref{eqtn1})$ the required nonvanishing components of ${\cal S}^{ijkl}$ are 
\begin{eqnarray}
{\cal S}^{0101} &=&-e^{2m+2n}\text{,}  \nonumber \\
{\cal S}^{0110} &=&e^{2m+2n} \text{,}  \nonumber \\
{\cal S}^{0202} &=&-e^{2l+2n}\text{,}  \nonumber \\
{\cal S}^{0220} &=&e^{2l+2n} \text{,}  \nonumber \\
{\cal S}^{0303} &=&-e^{2l+2m} \text{,}  \nonumber \\
{\cal S}^{0330} &=&e^{2l+2m}\text{.}  \label{eqtn9}
\end{eqnarray}
Using the above results in $(\ref{eqtn4})$ and $(\ref{eqtn5})$ we obtain 
\begin{equation}
L^{00}=L^{\alpha 0}=0 \text{.}  \label{eqtn10}
\end{equation}

\section{The Energy-momentum complex of Papapetrou}

The symmetric energy-momentum complex of Papapetrou (1948) is given by 
\begin{equation}
\Omega ^{ij}=\frac{1}{16\pi }{\cal {N}}_{\quad ,kl}^{ijkl}  \label{eqtn11}
\end{equation}
where 
\begin{equation}
{\cal {N}}^{ijkl}=\sqrt{-g}\left( g^{ij}\eta ^{kl}-g^{ik}\eta
^{jl}+g^{kl}\eta ^{ij}-g^{jl}\eta ^{ik}\right)  \label{eqtn12}
\end{equation}
and 
\begin{equation}
\eta ^{ik}=\left( 
\begin{array}{cccc}
1 & 0 & 0 & 0 \\ 
0 & -1 & 0 & 0 \\ 
0 & 0 & -1 & 0 \\ 
0 & 0 & 0 & -1
\end{array}
\right)  \label{eqtn13}
\end{equation}
is the Minkowski metric. The energy-momentum complex of Papapetrou satisfies
the local conservation laws 
\begin{equation}
\frac{\partial \Omega ^{ik}}{\partial x^{k}}=0\text{.}  \label{eqtn14}
\end{equation}
The locally conserved energy-momentum complex $\Omega ^{ik}$ contains
contributions from the matter, non-gravitational and gravitational fields. $%
\Omega ^{00}$ \ and $\Omega ^{\alpha 0}$ \ are the energy and momentum
(energy current) density components. The energy and momentum are given by 
\begin{equation}
P^{i}=\int \int \int \Omega ^{i0}dx^{1}dx^{2}dx^{3}  \label{eqtn15}
\end{equation}
We wish to find the energy and momentum density components for the
space-time described by the line element $(\ref{eqtn1})$. The required
nonvanishing components of ${\cal {N}}^{ijkl}$ are

\begin{eqnarray}
{\cal {N}}^{0011} &=&-(1+e^{-2l})e^{l+m+n}\text{,}  \nonumber \\
{\cal {N}}^{0110} &=&e^{-l+m+n}\text{,}  \nonumber \\
{\cal {N}}^{0022} &=&-(1+e^{-2m})e^{l+m+n}\text{,}  \nonumber \\
{\cal {N}}^{0220} &=&e^{l-m+n}\text{,}  \nonumber \\
{\cal {N}}^{0033} &=&-(1+e^{-2n})e^{l+m+n}\text{,}  \nonumber \\
{\cal {N}}^{0330} &=&e^{l+m-n}\text{,}  \nonumber \\
{\cal {N}}^{0101} &=&{\cal {N}}^{0202}={\cal {N}}^{0303}=e^{l+m+n}\text{.}
\label{eqtn16}
\end{eqnarray}
Using the above results in $(\ref{eqtn11})$ and $(\ref{eqtn12})$ we obtain 
\begin{equation}
\Omega^{00}=\Omega^{\alpha 0}=0 \text{.}  \label{eqtn17}
\end{equation}


\section{The Weinberg energy-momentum complex}

The symmetric energy-momentum complex of Weinberg (1972) is given by 
\begin{equation}
W^{ij}=\frac{1}{16\pi }{\Delta }_{\quad ,i}^{ijk}  \label{eqtn18}
\end{equation}
where 
\begin{equation}
{\Delta }^{ijk}=\frac{\partial h_{a}^{a}}{\partial x_{i}}\eta ^{jk}-\frac{%
\partial h_{a}^{a}}{\partial x_{j}}\eta ^{ik}-\frac{\partial h^{ai}}{%
\partial x^{a}}\eta ^{jk}+\frac{\partial h^{aj}}{\partial x^{a}}\eta ^{ik}+%
\frac{\partial h^{ik}}{\partial x_{j}}-\frac{\partial h^{jk}}{\partial x_{i}}
\label{eqtn19}
\end{equation}
and 
\begin{equation}
h_{ij}=g_{ij}-\eta _{ij}\text{.}  \label{eqtn20}
\end{equation}
$\eta _{ij}$ \ is the Minkowski metric (see Eq. $(\ref{eqtn13})$). The
indices on\ $\ h_{ij}$ or \ $\frac{\partial }{\partial x_{i}}$ are raised or
lowered with the help of $\eta $'s. The Weinberg energy-momentum complex $%
W^{ik}$ contains contributions from the matter, non-gravitational and
gravitational fields, and satisfies the local conservation laws 
\begin{equation}
\frac{\partial W^{ik}}{\partial x^{k}}=0\text{.}  \label{eqtn21}
\end{equation}
$\ W^{00}$ \ and \ $W^{\alpha 0}$ \ are the energy and momentum density
components. \ The energy and momentum \ components are given by 
\begin{equation}
P^{i}=\int \int \int W^{i0}dx^{1}dx^{2}dx^{3}  \label{eqtn22}
\end{equation}
We are interested in determining the energy and momentum density components.
Now using the equations $(\ref{eqtn1})$ and $(\ref{eqtn19})$ we find that
all the components of ${\Delta }^{ijk}$ vanish. Thus Eq. $(\ref{eqtn18})$
yields 
\begin{equation}
W^{ik}=0.  \label{eqtn23}
\end{equation}

\section{ Discussion and Summary}

The subject of energy-momentum localization in a curved space-time has been
controversial. Misner {\em et al.} (1973) argued that the energy is
localizable only for spherical systems. Cooperstock and Sarracino (1978)
contradicted their viewpoints and argued that if the energy is localizable
in spherical systems then it is localizable for all systems. Bondi (1990)
advocated that a nonlocalizable form of energy is not admissible in
relativity; therefore its location can in principle be found. The
energy-momentum complexes are nontensorial under general coordinate
transformations and are restricted to computations in ``Cartesian''
coordinates only. There has been a ``folklore'' that different
energy-momentum complexes are very likely to give different and hence
unacceptable energy distributions in a given space-time. To this end
Virbhadra and coworkers and some others showed that several energy-momentum
complexes ``coincide'' and give acceptable results for some well-known
space-times. Their results fascinated many researchers to work on this
subject.

In recent years some researchers showed interest in studying the energy
content of the universe in different models (see Rosen 1994, Cooperstock
1994, Johri {\em et al.} 1995, Banerjee and Sen 1997). Rosen (1994), with
the Einstein energy-momentum complex, studied the total energy of a closed
homogeneous isotropic universe described by the Friedmann-Robertson-Walker
(FRW) metric and found that to be zero. Using the Landau and Lifshitz
definition of energy Johri {\em et al.} (1995) demonstrated that (a) the
total energy of an FRW spatially closed universe is zero at all times
irrespective of equations of state of the cosmic fluid and (b) the total
energy enclosed within any finite volume of the spatially flat FRW universe
is zero at all times. Banerjee and Sen (1997) showed that the energy and
momentum density components vanish in the Bianchi type I space-times (they
used the energy-momentum complex of Einstein).

It is usually suspected that different energy-momentum complexes could give
different results for a given geometry. Therefore, we extended the
investigations of Banerjee and Sen with three more energy-momentum complexes
(proposed by Landau and Lifshitz, Papapetrou, and Weinberg) and found the
same results (see equations $(\ref{eqtn10})$, $(\ref{eqtn17})$ and $(\ref
{eqtn23})$) as reported by them. Note that the energy density component of
the energy-momentum tensor is not zero for the Bianchi type I solutions (see
Eq. $(\ref{eqtn2})$); however, it is clear from equations $(\ref{eqtn10})$, $%
(\ref{eqtn17})$ and $(\ref{eqtn23})$ that the total energy density (due to
matter plus field, as given by the energy-momentum complexes) vanishes
everywhere . This is because the energy contributions from the matter and
field inside an arbitrary two-surface in Bianchi type I space-times cancel
each other. The results in this paper advocate the importance of
energy-momentum complexes (opposes the ``folklore'' against them that
different complexes could give different meaningless results for a given
metric) and also supports the viewpoint of Tryon.

\acknowledgments
I am grateful to K. S. Virbhadra for guidance, to G. F. R. Ellis for
hospitality at the university of Cape Town, and to NRF for financial support.


\end{document}